\documentclass[fleqn,usenatbib]{mnras}

\usepackage{newtxtext,newtxmath}

\usepackage[T1]{fontenc}

\DeclareRobustCommand{\VAN}[3]{#2}
\let\VANthebibliography\thebibliography
\def\thebibliography{\DeclareRobustCommand{\VAN}[3]{##3}\VANthebibliography}

\usepackage{acronym}
\usepackage{graphicx}	
\usepackage{xcolor}
\usepackage{ulem} 

\newcommand{\new}[1]{\textbf{#1}}

\def\Mo{{\rm M_{\odot}}}

\title[What if GW190425 did not produce a BH promptly?]{%
What if GW190425 did not produce a black hole promptly?}

\author[D.~Radice et al.]{
David Radice$^{1,2,3}$\thanks{E-mail: david.radice@psu.edu}\thanks{Alfred P.~Sloan fellow},
Giacomo Ricigliano$^{4}$,
Mukul Bhattacharya$^{1,2}$,
Albino Perego$^{5,6}$,
\newauthor\ Farrukh J. Fattoyev$^{7}$,
and Kohta Murase$^{1,2,3,8}$
\\
$^{1}$ Institute for Gravitation \& the Cosmos, The Pennsylvania State University, University Park PA 16802, USA\\
$^{2}$ Department of Physics, The Pennsylvania State University, University Park PA 16802, USA\\
$^{3}$ Department of Astronomy \& Astrophysics, The Pennsylvania State University, University Park PA 16802, USA\\
$^{4}$ Technische Universit{\"a}t Darmstadt, Institut f{\"u}r Kernphysik, Schlossgartenstr. 2, 64289 Darmstadt, Germany \\
$^{5}$ Dipartimento di Fisica, Universit\`{a} di Trento, Via Sommarive 14,
  38123 Trento, Italy \\
$^{6}$ INFN-TIFPA, Trento Institute for Fundamental Physics and Applications, via Sommarive 14, I-38123 Trento, Italy \\
$^{7}$ Department of Physics and Astronomy, Manhattan College, Riverdale, NY 10471, USA\\
$^{8}$ Center for Gravitational Physics and Quantum Information, Yukawa Institute for Theoretical Physics, Kyoto University, Kyoto, Kyoto 606-8502, Japan
}

\date{Accepted XXX. Received YYY; in original form ZZZ}

\pubyear{2023}

\begin{document}
\label{firstpage}
\pagerange{\pageref{firstpage}--\pageref{lastpage}}
\maketitle

\begin{abstract}
It is widely believed that the binary neutron star merger GW190425
produced a black hole promptly upon merger. Motivated by the potential
association with the fast radio burst FRB 20190425A, which took place
2.5~hours after the merger, we revisit the question of the outcome of
GW190425 by means of numerical relativity simulations. We show that
current laboratory and astrophysical constraints on the equation of
state of dense matter do not rule out the formation of a long-lived
remnant. However, the formation of a stable remnant would have produced
a bright kilonova, in tension with upper limits by ZTF at the location
and time of FRB 20190425A. Moreover, the ejecta would have been
optically thick to radio emission for days to months, preventing a
putative FRB from propagating out. The predicted dispersion measure
is also several orders of magnitude larger than that observed for FRB
20190425A. Our results indicate that FRB 20190425A and GW190425 are not
associated.  However, we cannot completely rule out the formation of a
long-lived remnant, due to the incomplete coverage of the relevant sky
regions.  More observations of GW190425-like events, including potential
upper limit, have the potential to constrain nuclear physics. To this
aim, it is important that follow-up observational campaigns of
gravitational wave events are informed by the properties of the source,
such as their chirp mass, and we urge the LIGO-Virgo-KAGRA collaboration to
promptly release them publicly.
\end{abstract}



\section{Introduction}
\label{sec:introduction}
GW190425 \citep{Abbott:2020uma, LIGOScientific:2021djp} was the second
\ac{NS} merger detected by the Advanced LIGO
\citep{TheLIGOScientific:2014jea} and Virgo \citep{TheVirgo:2014hva}
detectors. It had a chirp mass $\mathcal{M} = 1.44 \pm 0.02\ \Mo$ and a
total mass $M = 3.4_{-0.1}^{+0.3}\ \Mo$, both more than $5\sigma$ larger
than the mean for galactic binary \ac{NS} systems. It is not excluded
that GW190425 was the merger of a light \ac{NS} and a low-mass black
hole (BH; \citealt{Han:2020qmn})\acused{BH}, although this might be in
mild tension with the lack of a kilonova associated with this event
\citep{Kyutoku:2020xka, Coughlin:2019xfb, Antier:2020nuy,
GravityCollective:2021kyg, Rastinejad:2021cko}.Assuming that both
objects were slowly spinning restricts the mass ratio and the total
mass to $q = 1{-}1.25$ and $M = 3.2{-}3.4\ \Mo$
\citep{Abbott:2020uma}. These values are compatible with the
hypothesis that there were two \acp{NS} in GW190425. Assuming that both
objects in the binary were \acp{NS}, \citet{Foley:2020kus} argued that
the total mass of GW190425 can be restricted even further to $M =
3.33_{-0.06}^{+0.10}\ \Mo$ and the mass ratio $q = M_1/M_2 =
1.27^{+0.40}_{-0.27}$.

It is commonly believed that GW190425 produced a \ac{BH} promptly after
merger. Indeed, \citet{Abbott:2020uma} estimate that the remnant
underwent prompt collapse with 96\% probability with the method proposed
by \cite{Agathos:2019sah}. Furthermore, numerical relativity simulations
targeted to GW190425 \citep{Dudi:2021abi, Camilletti:2022jms} found
prompt \ac{BH} formation in all considered cases. Perhaps unexpectedly
\citep{Barbieri:2020ebt, Raaijmakers:2021slr}, simulations, particularly
those with realistic microphysics \citep{Camilletti:2022jms}, predicted
small ejecta masses and faint kilonova for most configurations. As such,
the observational upper limits turned out to be only weakly constraining
on the binary parameters. It should also be remarked that not all of the
plausible sky locations for GW190425, which spanned ${\sim} 10^4~{\rm
deg}^2$ of the celestial sphere, have been observed.

A possible \ac{SGRB} coincident in time with GW190425 was identified
\citep{Pozanenko:2019lwh}, but further analysis revealed that its likely
origin is a magnetar flare in a nearby galaxy (NGC 253;
\citealt{Minaev:2020dce}). Searches for \ac{SGRB} afterglow emission in
GW190425 also did not identify any candidate \citep{Boersma:2021gyq}.

\citet{Moroianu:2022ccs} discussed the possible association between
GW190425 and the fast radio burst (FRB) 20190425A. FRBs are millisecond
duration pulses of bright radio emission that are located at
cosmological distances \citep{Lorimer:2007qn, Thornton:2013iua,
Petroff:2021wug}. Although a substantial sub-population of repeating
FRBs has been detected (see e.g., \citealt{Spitler:2016dmz,
Fonseca:2020cdd, Bhardwaj:2021xaa, Niu:2021bnl}), cataclysmic origins
for the larger sample of FRBs that do not repeat has not been ruled out.
Among many FRB source models within the realm of such cataclysmic
events, BNS mergers remain as a potential source of radio emission at
different stages of evolution of the central remnant (see
\citealt{Platts:2018hiy}). 
FRB 20190425A occurred 2.5~hours after the merger in the \ac{GW} sky
localization area. \citet{Moroianu:2022ccs} estimate the probability of
a chance coincidence to be 0.0052 (corresponding to 2.8~$\sigma$). A
possible interpretation of this signal can be made in the context of the
``blitzar'' mechanism of FRB \citep{Falcke:2013xpa, Most:2018abt}.
Accordingly, coherent radio emission can be launched by the snapping of
the field lines of a \ac{NS} when it suddenly collapses to \ac{BH}. Such
interpretation would imply not only that GW190425 did not promptly
produce a \ac{BH}, but also that the remnant survived for 2.5~hours
before collapsing to \ac{BH}, which would have profound implications for
the physics of dense matter \citep{Tews:2020ylw, Lim:2020zvx,
Fattoyev:2020cws, Godzieba:2020tjn}.  On the other hand,
\citet{Bhardwaj:2023avo} argued that the radio waves produced by the
collapse of the remnant in GW190425 would not be able to propagate
through the merger ejecta. This would exclude an association between
GW190425 and FRB 20190425A, unless the merger produced a very small
amount of ejecta.

In this paper, we revisit the question of the fate of GW190425. We
perform numerical relativity \ac{NS} merger simulations targeted to
GW190425 using the ``Big Apple'' (BA) relativistic mean-field theory
\ac{EOS} \citep{Fattoyev:2020cws}.  This \ac{EOS} satisfies many extant
laboratory and astronomical constraints, yet predicts the formation of
\textit{dynamically stable} \ac{RMNS} remnants for GW190425, which could
conceivably be supported over a timescale of hours after the merger.
However, we show that such scenario produces a bright kilonova that
would have been detected at the distance and location of FRB 20190425A.
Moreover, our simulations also show that the column density of the
ejecta emitted during the formation and the evolution of long-lived
remnants would have been too dense to be traversable by a putative FRB
pulse. Our simulations thus exclude an association between FRB 20190425A
and GW190425. They also disfavor the BA \ac{EOS}, with the caveat that,
since not all the sky region associated with GW190425 was observed,
GW190425 might have produced a bright kilonova that escaped detection.
All in all, our work makes the case for continued effort to follow up
GW190425-like events electromagnetically.

The rest of this paper is organized as follows.  In
Sec.~\ref{sec:simulations} we present the merger simulation results.
The associated kilonova signal is discussed in Sec.~\ref{sec:kilonova},
while the properties of a potential FRB are discussed in
Sec.~\ref{sec:frb}. Finally, Sec.~\ref{sec:conclusions} is dedicated to
discussion and conclusions.

\section{Merger Simulations}
\label{sec:simulations}

\begin{table*}
\caption{
    Mass of the each binary component $M_1$ and $M_2$, total binary baryonic
    mass $M_b$, reduced tidal parameter $\tilde\Lambda$, disk and ejecta masses
    $M_{\rm disk}$ and $M_{\rm ej}$ at the end of the simulations, and \new{root-mean-squared}
    opening angle $\sqrt{\langle (90^\circ - \theta)^2 \rangle_{\rm ej}}$ and velocity
    $\sqrt{\langle v^2 \rangle_{\rm ej}}$ of the ejecta for each binary.
}
\begin{center}
\label{tab:models}
\begin{tabular}{lcccccccc}
\hline\hline
Model & $M_1$       & $M_2$       & $M_b$       & $\tilde\Lambda$ & $M_{\rm disk}$ & $M_{\rm ej}$ & $\sqrt{\langle (90^\circ - \theta)^2 \rangle_{\rm ej}}$ & $\sqrt{\langle v^2 \rangle_{\rm ej}}$ \\
      & $[M_\odot]$ & $[M_\odot]$ & $[M_\odot]$ &                 & $[M_\odot]$    & $[M_\odot]$ & [deg] & $[c]$ \\
\hline
\texttt{q1.00-LR} & $1.654$ & $1.654$ & $3.700$ & $286$ & $0.28$ & $0.0120$ & $34$ & $0.23$ \\
\texttt{q1.00-SR} & $1.654$ & $1.654$ & $3.700$ & $286$ & $0.18$ & $0.0071$ & $34$ & $0.22$ \\
\texttt{q1.17-LR} & $1.795$ & $1.527$ & $3.721$ & $283$ & $0.26$ & $0.0054$ & $33$ & $0.23$ \\
\texttt{q1.17-SR} & $1.795$ & $1.527$ & $3.721$ & $283$ & $0.20$ & $0.0049$ & $32$ & $0.26$ \\
\texttt{q1.33-LR} & $1.914$ & $1.437$ & $3.766$ & $278$ & $0.25$ & $0.0128$ & $25$ & $0.22$ \\
\texttt{q1.33-SR} & $1.914$ & $1.437$ & $3.766$ & $278$ & $0.25$ & $0.0136$ & $29$ & $0.22$ \\
\texttt{q1.67-SR} & $2.149$ & $1.289$ & $3.904$ & $217$ & $0.12$ & $0.0070$ & $13$ & $0.22$ \\
\hline\hline
\end{tabular}
\end{center}
\end{table*}

For our simulations we use the BA \ac{EOS} \citep{Fattoyev:2020cws},
which predicts maximum nonrotating \ac{NS} mass of $2.6\ \Mo$ and radius
of a nonrotating $1.4\ \Mo$ \ac{NS} of $R_{1.4} = 12.96\ {\rm km}$.  We
construct irrotational initial data with the \texttt{LORENE}
pseudo-spectral code \citep{Gourgoulhon:2000nn} at the initial
separation of 45~km, corresponding to the last ${\sim}4$ orbits of the
binary. We fix the chirp mass of the system to $\mathcal{M} = 1.44\ \Mo$
and consider four mass ratios $q = 1, 1.17, 1.33$ and $1.67$.  The
properties of the initial data are summarized in Tab.~\ref{tab:models}.

We evolve the initial data using the general-relativistic hydrodynamics
code \texttt{WhiskyTHC} \citep{Radice:2012cu, Radice:2013hxh,
Radice:2013xpa, Radice:2015nva}, which is based on the \texttt{Einstein
Toolkit} \citep{Loffler:2011ay, EinsteinToolkit:2021_05}. For the
simulations discussed here, we use the \texttt{Carpet} \ac{AMR} driver
\citep{Schnetter:2003rb, Reisswig:2012nc}, which implements the
Berger-Oilger scheme with refluxing \citep{Berger:1984zza,
1989JCoPh..82...64B}. All simulations are performed at two resolutions:
LR (corresponding to $\Delta x_{\rm min} = 0.167  G M_\odot / c^2
 \simeq  246\ {\rm m}$) and SR ($\Delta x_{\rm min} = 0.125 G M_\odot /
c^2 \simeq 185\ {\rm m}$). However, we do not report the results of the
\texttt{q1.67-LR} simulation, since it failed shortly after \ac{BH}
formation. We evolve the spacetime geometry using the \texttt{CTGamma}
code \citep{Pollney:2009yz, Reisswig:2013sqa}, which solves the Z4c
formulation of Einstein's equations \citep{Bernuzzi:2009ex,
Hilditch:2012fp}. The standard first-order moving puncture gauge is
employed \citep{vanMeter:2006vi}. The zero-temperature BA \ac{EOS} is
augmented with a ideal-gas thermal component with adiabatic index
$\Gamma_{\rm th} = 1.7$, following \textit{e.g.}
\citet[][]{Shibata:2005ss, Bauswein:2010dn, Endrizzi:2018uwl,
Figura:2020fkj}.

\begin{figure}
  \includegraphics[width=\columnwidth]{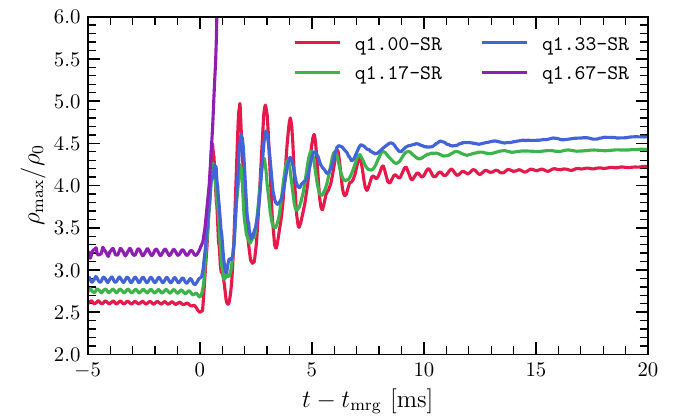}
  \caption{Maximum density for all \texttt{SR} binaries as a function of
  time. With the exception of the \texttt{q1.67-SR} binary, all models
  produce stable remnants.}
  \label{fig:rho.max}
\end{figure}

The BA EOS not only predicts a $2.6\ \Mo$ non-rotating maximum NS but
also agrees with a host of both nuclear and astrophysical observables.
The parameters of this model are tuned to reproduce the ground-state
properties of finite nuclei, such as binding energies and charge radii.
The BA EOS also successfully predicts the bulk parameters of nuclear
matter in agreement with bulk parameters constrained from both the giant
monopole resonances and neutron skins.  The soft density dependence of
symmetry energy with $L = 39.8$ MeV allows the model to predict the
$1.4\ \Mo$ NS with radius of 12.96~km and dimensionless tidal
polarizability of $717.3$, consistent with astrophysical observations
\citep{Riley:2019yda, Miller:2019cac, LIGOScientific:2018cki,
Breschi:2021tbm}.  The required stiffening of the BA EOS at high
densities allows the model to predict a maximum non-rotating NS mass
of $2.6\ \Mo$, which is in mild tension with the existing constraints
obtained from energetic heavy-ion collisions \citep{Danielewicz:2002pu},
but it is still viable within $2\sigma$. In fact, even more extreme
models can be produced achieving maximum nonrotating \ac{NS} masses in
excess of $2.8\ \Mo$ as shown by \citet{Mueller:1996pm,
Fattoyev:2020cws}.

The maximum baryonic mass for a uniformly rotating \ac{NS} predicted by
the BA \ac{EOS} is $3.834\ \Mo$.  As can be seen in
Tab.~\ref{tab:models}, all binaries with the exception of $q = 1.67$,
have baryonic mass below this value. As such, they are expected to form
supermassive remnants that are stable on secular timescales
\citep{1992ApJ...398..203C, Baumgarte:1999cq, Radice:2018xqa,
Radice:2023zlw}. This expectation is confirmed by the simulations. In
Fig.~\ref{fig:rho.max} we plot the maximum density for all SR
simulations. After a period of violent oscillations, the merger remnants
for the $q = 1$, $q = 1.17$ and $q = 1.33$ settle to a quasi-steady
configuration comprising a massive \ac{RMNS} surrounded by a thick
accretion disk. In contrast, the $q = 1.67$ binary undergoes prompt
\ac{BH} formation upon merger.

\begin{figure*}
  \includegraphics[width=\textwidth]{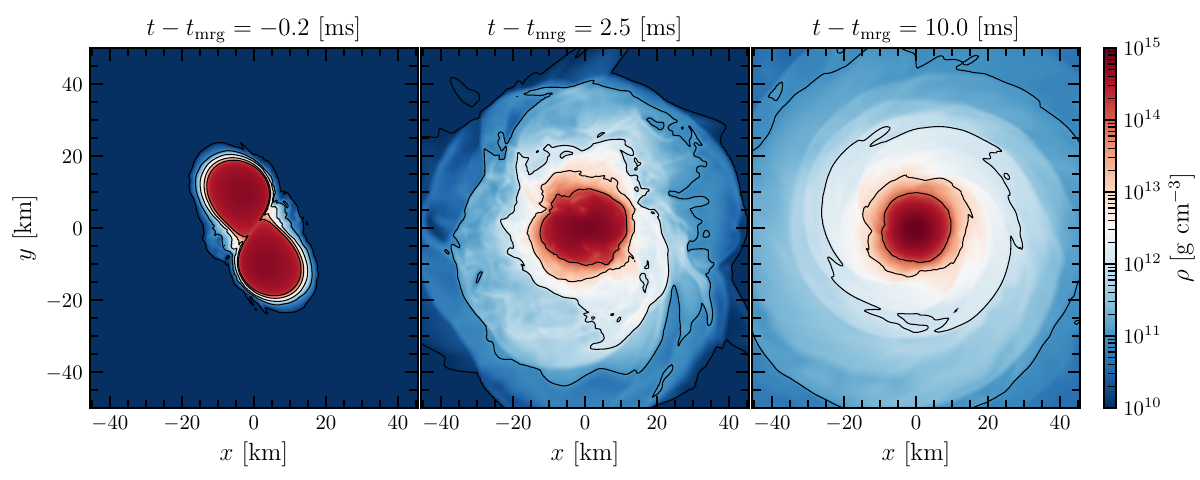}
  \caption{Rest-mass density on the orbital plane for the
  \texttt{q1.00-SR} binary. The merger results in the formation of a
  long-lived remnant.}
  \label{fig:q1.xy}
\end{figure*}

\begin{figure*}
  \includegraphics[width=\textwidth]{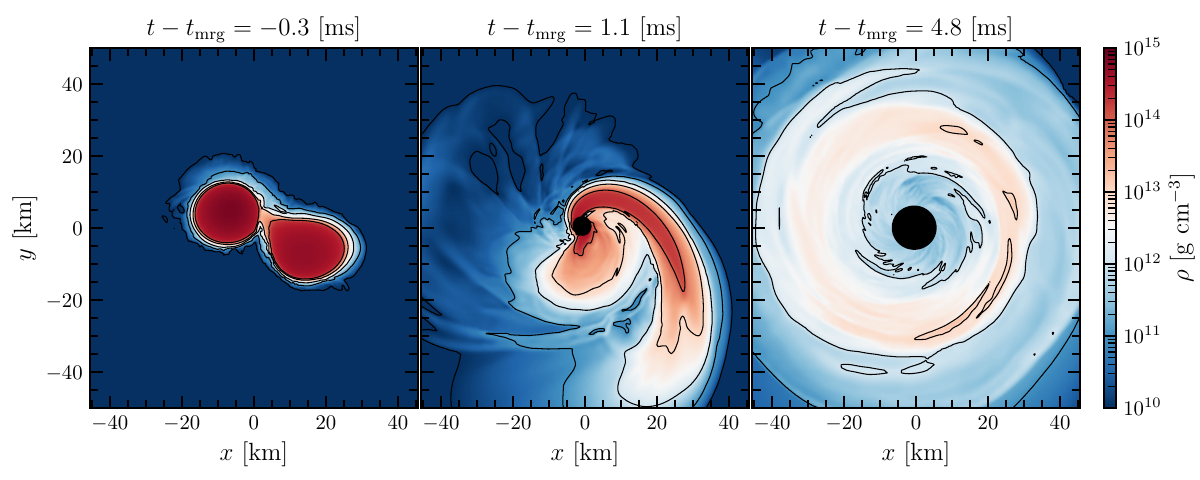}
  \caption{Rest-mass density on the orbital plane for the
  \texttt{q1.67-SR} binary. The black contour encloses the region with
  $\alpha < 0.2$, which is expected to be located within the apparent
  horizon. During the merger the secondary star is disrupted and the
  primary collapses to BH. The outcome of the merger is a BH surrounded
  by a thick accretion disk.}
  \label{fig:q167.xy}
\end{figure*}

Figure~\ref{fig:q1.xy} shows the dynamics of the material in the orbital
plane for the \texttt{q1.00-SR} binary, which is representative of all
the stable remnant cases. This should be contrasted with
Fig.~\ref{fig:q167.xy} which shows the dynamics for the
\texttt{q1.67-SR}. In this case, the secondary \ac{NS} is tidally
disrupted during its last orbit. The primary \ac{NS} collapses to form a
\ac{BH} as a result of the accretion from the tidal stream. A similar
dynamics has also been reported in \citet{Bernuzzi:2020txg} for a
lower mass system with a different \ac{EOS}.

\begin{figure*}
  \includegraphics[width=0.98\columnwidth]{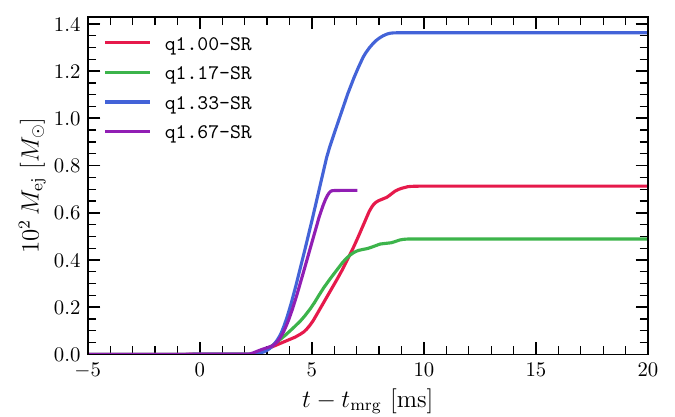}
  \hfill
  \includegraphics[width=0.98\columnwidth]{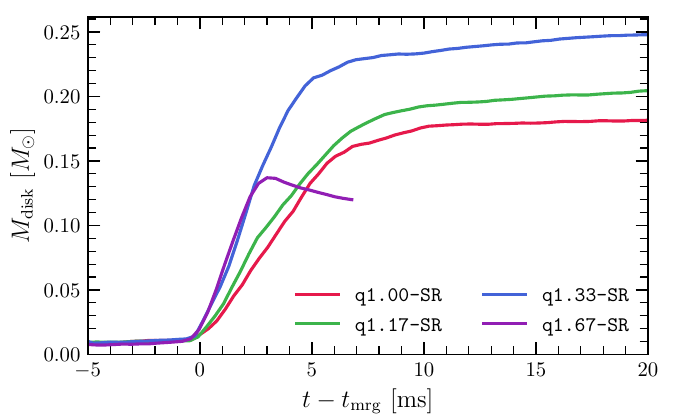}
  \caption{Dynamical ejecta mass (\emph{left panel}) and remnant disk
  mass (\emph{right panel}) as a function of time for all the
  \texttt{SR} models. All systems produce massive outflows and result in
  the formation of very massive disks.}
  \label{fig:ejecta}
\end{figure*}

All binaries produce massive ejecta, driven by tidal torques and shocks
during the merger, including the prompt collapse $q = 1.67$ binary. We
estimate the mass ejection by integrating the flux of gravitationally
unbound matter (with $u_t < -1$, $u$ being the fluid four-velocity)
crossing a coordinate sphere of radius $200\, G/c^2\, M_\odot \simeq
295\ {\rm km}$. The results are reported in Tab.~\ref{tab:models} and
Fig.~\ref{fig:ejecta} (left panel). The dynamical ejecta mass ranges
from $0.49 \times 10^{-2}\ \Mo$ to $1.36 \times 10^{-2}\ \Mo$. Because
dynamical ejecta are driven by a combination of shock heating, which
weakens as the mass ratio increases, and tidal torques, which has the
opposite mass ratio dependency, the dynamical ejecta mass is not
monotonic with $q$ \citep{Dietrich:2016hky}.  Finite-resolution errors
are larger for the $q = 1$ binary, since shocks are more challenging to
capture numerically. That said, we caution the reader that our
simulations might be overestimating the ejecta mass, since we are
neglecting finite-temperature and neutrino effects
\citep{Nedora:2020qtd}.  However, the resulting uncertainty is
subdominant compared to that in the mass of the secular ejecta
from the merger disk.

All binaries also result in the formation of accretion disks. To
quantify this, we compute the total mass of material with rest-mass
density $\rho < 10^{13}\ {\rm g}\ {\rm cm}^{-3}$, which we assume to be
belonging to the disk, following \citet{Shibata:2017xdx} and
\citet{Radice:2018pdn}. This is motivated by the fact that this density
roughly corresponds to the transition to a quasi-Keplerian rotational
profile in the remnant \citep{Hanauske:2016gia}. For the $q = 1.67$
binary, we also exclude material with lapse function below $0.2$, which
would be located inside the \ac{BH} \citep{Bernuzzi:2020txg}. The
results are reported in Tab.~\ref{tab:models} and Fig.~\ref{fig:ejecta}
(right panel). We find that all binaries form very massive disks, up to
$0.28\ \Mo$. It is expected that $20{-}50\%$ of the disks will evaporate
due to neutrino-driven winds, magnetic torques, and nuclear
recombination \citep{Metzger:2014ila, Radice:2018xqa, Fahlman:2018llv,
Nedora:2019jhl, Fujibayashi:2020dvr, Just:2023wtj}. As such, this
secular component of the ejecta is expected to be dominant. We remark
that the disk is still rapidly accreting onto the \ac{BH} for the
\texttt{q1.67-SR} binary at the time when we stop the simulation. This
means that the disk mass we report for \texttt{q1.67-SR} should be taken
as an upper bound. That said, because this binary underwent prompt
collapse, it is not a suitable candidate for FRB 20190425A. For this
reason, we will focus on the $q = 1$, $q = 1.17$, and $q = 1.33$
binaries for the rest of the discussion.

\section{Kilonova Light Curves}
\label{sec:kilonova}

\begin{figure*}
  \includegraphics[width=0.98\textwidth]{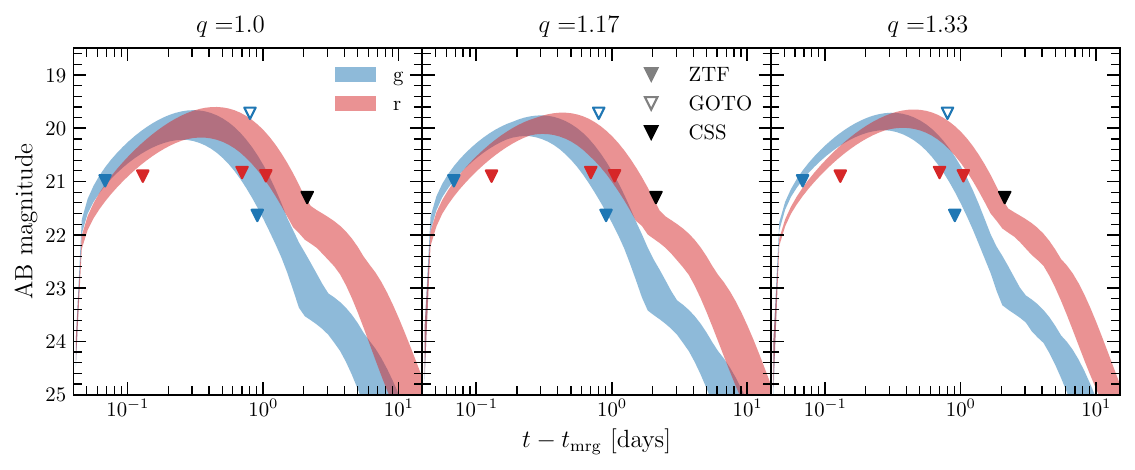}
  \caption{Synthetic kilonova color curves in the $r$ and $g$ bands, for
  a source at a distance $D_L\simeq142$ Mpc and inclination angle
  $\theta_{\rm view}\simeq56^\circ$, as obtained using as input the
  outcome of the $q = 1$, $q = 1.17$, and $q = 1.33$ simulations,
  together with observational upper limits from ZTF, GOTO and
  SAGUARO/CSS. The SAGUARO/CSS upper limit was taken without a filter
  and we compare it with the magnitudes computed both for the $g$ and
  $r$ filters.  The width of each band encodes the estimated
  uncertainties in the ejecta modeling and source properties (see the
  main text for more details).  The upper limits in the $r$ band from
  ZTF appears to rule out the formation of a stable \ac{RMNS} in
  GW190425, assuming to know the sky position as the one of FRB
  20190425A.}
  \label{fig:kn}
\end{figure*}

Based on the available ejecta properties extracted from the simulations, we
compute synthetic, broadband kilonova light curves during the first few
days after the merger.  We assume the kilonova emission to be powered by
the radioactive decay of freshly synthesized $r$-process elements in the
ejecta, while we neglect possibly
spin-down energy injection which would have resulted in a merger nova
that can be significantly brighter than kilonovae \citep{Yu:2013kra,
Metzger:2013cha, Kisaka:2015vma, Murase:2017snw, Sarin:2022wby,
Ai:2022pig, Wang:2023qww}.  Thus, we consider our prediction as a robust
lower limit for the electromagnetic counterpart.  We employ
\texttt{xkn}, a multi-component, anisotropic kilonova model, presented
in \citet{Ricigliano:2023svx} and based on the framework first described
and implemented in \citet{Martin:2015hxa} and \citet{Perego:2017wtu}.
In contrast to the original version, \texttt{xkn} computes the light
curves starting from a solution of the diffusion equation for photon
radiation emitted at the photosphere, obtained using a semi-analytical
approach first proposed by \citet{Wollaeger:2017ahm}. The model
employs a self-similar homologous density profile, which is expected to
well describe the ejecta profile at the time of the kilonova emission
\citep[see, e.g.,][]{PhysRevD.107.023016}. \citet{Ricigliano:2023svx}
extended this solution by including composition and time dependent
radioactive heating rates, thermalization efficiencies and opacities,
and by accounting for the contribution to the emission coming from the
external, optically thin layers of the ejecta.  \texttt{xkn}
requires as input of each component the total ejecta mass, the average
expansion velocity and the gray opacity, as well as the angular
distribution of these quantities.  Moreover, the model uses heating rates that are
parametrized in terms of the three quantities that determine the
nucleosynthesis in the homologously expanding ejecta, namely the
electron fraction, the specific entropy and the expansion timescale
\citep{Hoffman:1996aj}. This approach was also adopted in
\citet{Wu:2021ibi}\citep[see e.g.][for similar
fits.]{Lippuner:2015gwa, Setzer:2022nqs}

We consider two axisymmetric ejecta components for our kilonova model: a
faster, external dynamical component, and a slower, inner wind one.  The
angular profiles of the ejecta (mass fraction with angle,
\textbf{electron fraction}, and entropy) are only weakly dependent on
the \ac{EOS}, but are strongly impacted by the neutrino treatment used
in the simulations \citep{Radice:2016dwd, Perego:2017wtu}. Given that
our simulations do not implement neutrino transport, we take values
based on the results of neutrino-radiation simulations from the
literature. Specifically, regarding the dynamical ejecta, the values of
its total mass, $M_{\rm ej}$, and expansion velocity, $\sqrt{\langle v^2
\rangle}_{\rm ej}$, are the same as that of the simulated dynamical
ejecta, as reported in Table~\ref{tab:models}. Since the opening angle
of the expanding ejecta, $\sqrt{\langle \theta^2 \rangle}_{\rm ej}$, is
between 25 and 34 degrees, we consider a broad angular distribution in
the mass, characterized by a $\sin{\theta}$ dependence on the polar
angle $\theta$, measured from the orbital axis of the binary.  As a
consequence of the more intense neutrino irradiation inside the polar
regions, we expect an electron fraction $Y_e \gtrsim 0.25$ in this
region, so we set the gray photon opacity to $\kappa_{\rm dyn} = 1~{\rm
cm^2~g^{-1}}$ for $\theta \lesssim 45^{\circ}$.  In the region across
the equator, for $\theta \gtrsim 45^{\circ}$, we expect lower $Y_e$
and set $\kappa_{\rm dyn} = 15~{\rm cm^2~g^{-1}}$.  Such a value is
expected if a non-negligible amount of lanthanides are synthetized
inside the ejecta \citep[see, e.g.,][for typical grey opacity values of
r-process enriched material]{Tanaka:2019iqp}. We assume the average
velocity of the ejecta to be constant across different polar angles, and
we set its average specific entropy to $s=10\ k_{\rm B}\ {\rm
baryon^{-1}}$, while we compute the ejecta expansion timescale as
$\tau_{\rm exp}= 1~{\rm ms} \left(c/\sqrt{\langle v^2 \rangle_{\rm
ej}} \right)$.  For the second ejecta component, we cannot rely on
detailed ejecta properties extracted from our simulations, since the
latter are not long enough to model the disk evolution. We consider two
possibilities: a spherical secular wind, or an aspherical spiral wave
wind.  In both cases, we assume the ejecta mass to vary between 20\% and
40\% of the disk mass and we adopt again a $\sin{\theta}$ matter
distribution in the case of the aspherical spiral wave wind. In both
instances, we set a uniform entropy $s=20\ k_{\rm B}\ {\rm
baryon^{-1}}$, and an isotropic average velocity, with lower values
relative to the dynamical ejecta, i.e.  $\sqrt{\langle v^2 \rangle_{\rm
ej}}=0.06~c$ for the secular wind \citep[as expected from the nuclear
recombination of matter inside the disk and consistent with AT2017gfo
modelling, see e.g.][]{Villar:2017wcc,Perego:2017wtu}, and
$\sqrt{\langle v^2 \rangle_{\rm ej}}=0.15~c$ for the spiral wave wind
\citep[as obtained by detailed spiral wave wind simulations
in][]{Nedora:2019jhl,Nedora:2020qtd}. Due to the presence of a
\ac{RMNS}, the photon opacity of this ejecta component is expected to be
smaller than that of the equatorial part of the dynamical ejecta
\cite[see e.g.][]{Lippuner:2017bfm}.  Therefore, in the case of the
secular wind, we set the grey opacity to $\kappa_{\rm wind} = 5~{\rm
cm^2~g^{-1}}$, while for the spiral wave wind we assume $\kappa_{\rm
wind} = 1~{\rm cm^2~g^{-1}}$ for $\theta \lesssim 45^{\circ}$, and
$\kappa_{\rm wind} = 5~{\rm cm^2~g^{-1}}$ for $\theta \gtrsim
45^{\circ}$. Finally, we consider a source luminosity distance $D_L$
varying between $141.8$ and $142.7$ Mpc, correspondent to the value and
uncertainty in the redshift of the FRB host galaxy UGC10667
\citep{2009ApJS..182..543A}, together with an inclination angle
$\theta_{\rm view}$ between $46^\circ$ and $70^\circ$, as inferred by
\citet{Bhardwaj:2023avo}.

Figure~\ref{fig:kn} summarizes the results of the kilonova emission,
with the uncertainty in the amount of disk material expelled, the
structure of the disk wind component, the source distance and
inclination angle being represented by the width of the bands. The color
curves are compared to the observational upper limits in the $r$ and $g$
bands taken from \url{https://treasuremap.space/} \citep{Wyatt:2020qyw},
imposed by the Zwicky Transient Facility
\citep[ZTF,][]{2019PASP..131g8001G}, the Gravitational-wave Optical
Transient Observer \citep[GOTO;][]{2018cosp...42E2486O} and the Searches
After Gravitational waves Using ARizona Observatories
\citep[SAGUARO/CSS;][]{2012DPS....4421013C, Lundquist:2019cty,
Paterson:2020mmd}   \footnote{When our manuscript was in the final
editing stages, a new analysis of ATLAS and Pan-STARRS data was
announced \citep{Smartt:2023igh}. We do not include these data, but it
would not alter our conclusions.}. In the latter case, all images were
taken without a filter and were calibrated to Gaia $G$-band using Gaia
DR2 \citep{Gaia:2018ydn}. We compare the resulting upper limits for the
magnitudes computed both in the $g$ and $r$ band. While limits from GOTO
and CSS do not place any significant constraint on the model, we find
that the kilonova emission, especially in its redder component, appears
too bright to be compatible with the ZTF limits before and around 1 day,
regardless of the different simulated binary merger.  This is due to the
relatively high amount of material expelled by the disk during the
\ac{RMNS} lifetime, which exceeds by about one order of magnitude the
amount of dynamical ejecta.  We note that, in the context of the
kilonova model, the considered sources of uncertainty only place up to
one magnitude of variation on the overall emission. In order to
reconcile the model with the ZTF limits, we would need to invoke
considerably larger uncertainties. The latter can be potentially found
in the modeling of the nuclear heating and of the ejecta opacity.
Different nuclear inputs for the nucleosynthesis calculations lead to
variations of up to ${\sim} 1$ order of magnitude on the heating rate.
The opacities are even more uncertain, due to the incompleteness of
atomic data \citep[see][for a more detailed discussion of the
uncertainties of our kilonova model]{Ricigliano:2023svx}.  However, here
we absorb the ignorance on the ejecta opacity by considering different
ejecta setups.  We note that the heating rate uncertainty can be
translated in an additional ${\sim}0.25$~mag variation in our light
curves. Even so, our results would barely meet the ZTF constraints and,
as such, they tend to disfavor the formation of a long lived remnant at
the time of GW190425 and sky location of FRB 20190425A. In absence of an
association with FRB 20190425A, we can only exclude the formation of a
long-lived remnant if the distance to the source was $D_L \lesssim
150$~Mpc and if the merger happened in an area of the sky that has been
surveyed down to ${\sim}21$ magnitudes. On the other hand, the distance
inferred from the \ac{GW} alone is $159^{+69}_{-71}\ {\rm Mpc}$
\citep{Abbott:2020uma}, and therefore the current data are not strongly
constraining.

\section{Fast Radio Bursts}
\label{sec:frb}

A fast-rotating pulsar/magnetar that can be left as a merger remnant may
power not only FRB emission but also non-thermal emission
from pulsar/magnetar wind nebulae.  A significant fraction of
the spin-down and/or magnetic energy can be extracted by an outgoing
relativistic wind and carried by non-thermal electrons and positrons via
magnetic energy dissipation, leading to synchrotron and invese-Compton
emission. This scenario is widely considered to explain multi-wavelenth
spectra of Galactic pulsar wind nebulae \citep[][for a review]{Gaensler:2006ua}. 
Quasi-steady synchrotron emission from embryonic nebulae \citep{Murase:2016sqo}, 
can explain persistent radio counterparts of FRB 121102 and FRB 190520
\citep{Chatterjee:2017dqg, Niu:2021bnl}. Such synchrotron nebular emission
from a pulsar/magnetar has also been studied in the context of merger
remnants and the accretion-induced collapse of white dwarfs 
\citep{Kashiyama:2017ehl,Murase:2017snw,Yamasaki:2017hdr}.

The fact that FRB 20190425A occurred 2.5~hours after GW190425 motivates
us to consider FRB emission from a merger remnant in the post-merger
phase.  However, radio waves can be significantly depleted via various
absorption processes, including free-free absorption in the ejecta and
synchrotron absorption in the nebula \citep{Murase:2016ysq,
Bhardwaj:2023avo}.  The dense ejecta can also make a significant
contribution to the observed dispersion measure (DM), which can be used
as an important constraint on the model.

Assuming UGC 10667 at redshift $z=0.03136$ as the host for FRB 190425,
the DM inferred from the FRB can be written in terms of five primary
components, ${\rm DM}_{\rm obs} \simeq {\rm DM}_{\rm MW} + {\rm DM}_{\rm
halo} + {\rm DM}_{\rm IGM} + {\rm DM}_{\rm host}+{\rm DM}_{\rm src} =
128\,{\rm pc\, cm^{-3}}$, where ${\rm DM}_{\rm MW/halo}$ is the Milky
Way interstellar medium (ISM)/halo contribution, ${\rm DM}_{\rm IGM}$ is
the intergalactic medium (IGM) contribution and ${\rm DM}_{\rm host}$ is
the FRB host galaxy contribution including its halo. Here ${\rm DM}_{\rm
src}={\rm DM}_{\rm ej/w}+{\rm DM}_{\rm nb}$ includes contributions from
both the merger ejecta/wind and magnetized nebula. We mainly focus on
the merger ejecta, since their contribution is dominant. However, we
give an estimate for the optical depth of the pair nebula below.  From
theoretical and data-driven estimates of ${\rm DM}_{\rm MW}$, ${\rm
DM}_{\rm halo}$ and ${\rm DM}_{\rm IGM}$, using models for the average
electron number density along the FRB line-of-sight (see e.g.,
\citealt{Cordes:2002wz, Cordes:2003ik, Planck:2015fie,
2019MNRAS.485..648P, Platts:2020ynd}), we infer ${\rm DM}_{\rm
host}+{\rm DM}_{\rm src} \lesssim 30\,{\rm pc\,cm^{-3}}$.  Assuming a
negligible host galaxy contribution to ${\rm DM}_{\rm obs}$, we
numerically compute the number density of free electrons in the
ejecta/wind as $n_{\rm e,ej/w} \approx 3M_{\rm ej/w}/(4\pi R_{\rm
ej/w}^3 \mu_e m_{\rm H})$, to ascertain ${\rm DM}_{\rm src} \approx
n_{\rm e,ej}R_{\rm ej} + n_{\rm e,w}R_{\rm w}$.  Here $M_{\rm ej/w}$ is
the ejecta/wind mass, $R_{\rm ej/w}$ is the ejecta/wind radius and
$1/\mu_e$ corresponds to the number of free electrons per baryon for an
electron temperature of $\mathcal{T}_e \sim 10^4\, {\rm K}$
\citep{1984ApJ...287..282H}.

We compute the time evolution of DM, in which for the ejecta we assume a
spherically expanding distribution of mass until the NS collapse time
$t\simeq2.5\,{\rm hrs}$.  For the magnetar, we set the NS initial spin
period $P_i \sim 1\,{\rm ms}$, assuming that the remnant will achieve
solid body rotation with a spin close to the mass-shedding limit
\citep{Radice:2018xqa, Radice:2023zlw}.  The surface dipole field is
taken to be $B_{\rm dip}\sim 2\times10^{14}\,{\rm G}$, so that the
spindown timescale is consistent with the 2.5~hours delay between
GW190425 and FRB 20190425A. The interior magnetic field is $B_{\rm
int}\sim 10^{16}\,{\rm G}$ and is predominantly toroidal.  The radii of
the merger ejecta and magnetar-powered nebula are then calculated
following the merger nova model presented in \citet{Murase:2017snw}.
Due to a faster radius evolution for the ejecta and wind with an
increase in dipolar field strength $B_{\rm dip}$ (and for fixed $B_{\rm
int}\sim 10^{16}\,{\rm G}$), $DM_{\rm src} \approx n_{\rm e,ej}R_{\rm
ej} + n_{\rm e,w}R_{\rm w}$ decreases as the corresponding electron
number densities are smaller.
 
\begin{figure}
  \includegraphics[width=\columnwidth]{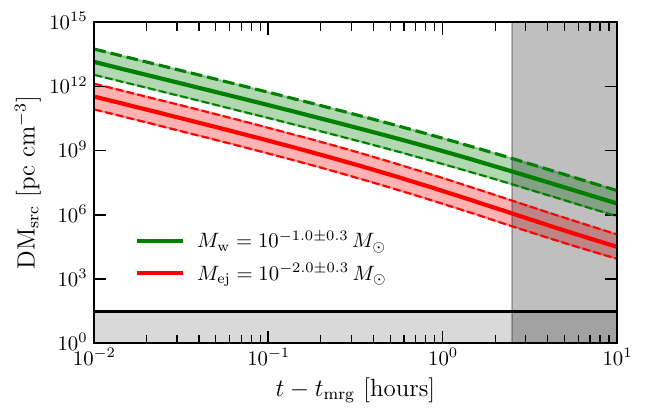}
  \caption{Time evolution of the DM contribution from dynamical ejecta
  (red) and disk wind (green) are shown. The representative ejecta
  masses are obtained from Tab.~\ref{tab:models} and
  Fig.~\ref{fig:ejecta}. The dashed lines for both components correspond
  to mass variation by a factor of two. The horizontal grey-shaded area
  bound by the solid black line shows the allowed region where ${\rm
  DM}_{\rm src} \lesssim 30\,{\rm pc\,cm^{-3}}$ (assuming ${\rm DM}_{\rm
  host}\approx 0$), as inferred from the radio observations. The
  vertical grey-shaded area ($t \gtrsim 2.5\,{\rm hr}$) denotes the
  region disallowed by the collapse of remnant NS to a BH.}
  \label{DM_tau_evol}
\end{figure}

The unshaded area in Fig.~\ref{DM_tau_evol} shows the time evolution of
${\rm DM}_{\rm src}$ prior to the collapse of the remnant NS to a BH. We
consider separately the contribution of dynamical ejecta and disk wind.
As expected, the dominant contribution to ${\rm DM}_{\rm src}$ arises
from the dense merger ejecta. We estimate the dynamical ejecta mass
($M_{\rm ej}$) and disk wind mass ($M_{\rm w}$) from the results
obtained in Tab.~\ref{tab:models} and Fig.~\ref{fig:ejecta}.  For the
dynamical ejecta, we consider a fiducial mass $M_{\rm
ej}=0.01\,M_{\odot}$, expansion velocity $v_{\rm ej}=0.25\,c$, energy
$\mathcal{E}_{\rm ej}=3\times10^{50}\,{\rm erg}$ and initial electron
fraction $Y_{\rm e,ej} \sim 0.15$ corresponding to mean atomic mass
number $\bar{A}\simeq 166$, average charge $\bar{Z}\simeq65$ and
$1/\mu_e\sim3.6/\bar{A}$ (at $2.5$~hours). The corresponding parameters
at $2.5$~hours for the disk wind are $M_{\rm w}=0.1\,M_{\odot}$, $v_{\rm
w}=0.1\,c$, $\mathcal{E}_{\rm w}=10^{51}\,{\rm erg}$, and $Y_{\rm e,w}
\sim 0.3$, resulting in $\bar{A}\simeq 81$, $\bar{Z}\simeq 34$ and
$1/\mu_e\sim2.4/\bar{A}$. The average atomic mass number and charges
have been obtained by performing single-trajectory nucleosynthesis
calculation with the \texttt{SkyNet} nuclear reaction network code
\citep{Lippuner:2017tyn}. The average ionization fraction is obtained
assuming local thermodynamic equilibrium.

We stress that our constraint from the observed DM is conservative.  It
should be noted that the assumption that the ionization fraction is
given by the Saha equation at 2.5~hours underestimates the ejecta DM and
opacity.  In particular, the ejecta is expected to be hotter and
potentially more ionized at early times.  X-ray and gamma-ray photons
may diffuse out from the nebula and increase the opacity in the radio
band \citep{Metzger:2013cha,Murase:2017snw,Wang:2023qww}. For our ejecta
parameters, we find that ${\rm DM}_{\rm src}$ significantly
overestimates the inferred near-source contribution ${\rm DM}_{\rm src}
\lesssim 30\,{\rm pc\,cm^{-3}}$, even when assuming ${\rm DM}_{\rm host}
\approx 0$.  

DM can also be larger due to free electron-positron pairs that can be
created in the magnetized nebula or NS magnetosphere through
electromagnetic cascades.  For a spin-down luminosity $L_{\rm
sd}\sim3\times{10}^{47}~{\rm erg}~{\rm s}^{-1}$ and nebular radius
$R_{\rm nb}\sim{10}^{14}~{\rm cm}$ at 2.5~hours, the compactness
parameter is estimated to be $l_e\sim L_{\rm sd}\sigma_T/(R_{\rm nb}
m_ec^3)\sim 8\times{10}^4$.  The resulting Thompson optical depth is
$\tau_T\approx n_{\pm,\rm nb}R_{\rm nb}\sigma_T \sim {({l_e {\rm
PY}})}^{1/2}\sim90$, where ${\rm PY} \sim 0.1$ is the pair yield
\citep{LZ87}, implying that the dispersion measure from the nebula,
${\rm DM}_{\rm nb}\approx n_{\pm,\rm nb}R_{\rm nb}$, would also exceed
${\rm DM}_{\rm obs}$.  Although we assume that the FRB emission is
produced near the RMNS surface at $R \sim 10^{6}\,{\rm cm}$, this does
not affect our constraints on ${\rm DM}_{\rm src} \lesssim 30\,{\rm
pc\,cm^{-3}}$. This is expected as most of the DM is accumulated either
in the merger ejecta with $R_{\rm ej} \sim 10^{13-14}\,{\rm cm}$ (baryon
dominated) or in the magnetized nebula (pair dominated) at $R_{\rm ej}
\sim 10^{11-13}\,{\rm cm}$, both with a significantly larger radius
compared to the remnant.

For FRB pulses that are produced at inner regions to be observed, both of the
ejecta and nebula should be optically thin to scattering or absorption
effects that can suppress the radio signal. The optical
depth for free-free absorption in the merger ejecta is given by \citep{Murase:2016ysq}
\begin{equation}
  \tau_{\rm ff}\approx
  2.1\times10^{-25}\, \mathcal{T}_{e,4}^{-1.35} \nu_{9}^{-2.1} \int {\rm d}r\,
  n_{\rm e,ej}\, n_{\rm i,ej}\, \bar{Z}^2\,,
\end{equation}
and should be sufficiently small. Here $n_{\rm i,ej}$ is the number
density of ions in the ejecta and $\bar{Z}$ is the average charge of the nuclei. 
Similarly, the optical depth for synchrotron absorption in the nebula,
\begin{equation}
  \tau_{\rm sa} = R_{\rm nb} \int {\rm d}\mathcal{E}_e\,
  \sigma_{\rm sa}(\nu, \mathcal{E}_e)
  \frac{{\rm d}n_{\mathcal{E}_e}}{{\rm d}\mathcal{E}_e}\,,
\end{equation}
should also not exceed unity for a given energy distribution of
electrons ${\rm d}n_{\mathcal{E}_e}/{\rm d}\mathcal{E}_e$ (see
\citealt{Murase:2016sqo} for more details) and an energy-dependent SSA
cross section $\sigma_{\rm sa}$ \citep{1991MNRAS.252..313G}.

For dynamical ejecta with $M_{\rm ej} \sim
0.01\,M_{\odot}$, the optical depths for free-free absorption $\tau_{\rm
ff} \sim 2.3\times10^{10}$ and synchrotron absorption $\tau_{\rm sa}
\sim 1.6\times10^{11}$ both significantly exceed unity at $t \simeq
2.5\,{\rm hrs}$, when the remnant NS undergoes collapse to a BH.
Similarly, for the disk wind with $M_{\rm w} \sim 0.1\,M_{\odot}$, the
corresponding optical depths are $\tau_{\rm ff} \sim 8.1\times10^{13}$
and $\tau_{\rm sa} \sim 7.1\times10^{10}$, respectively. These
demonstrate that both dynamical ejecta and disk winds are too dense at
$t \lesssim 2.5\,{\rm hrs}$ which results in significant attenuation of
the radio signal, thereby excluding an association between GW190425 and
FRB 190425.

Although our results qualitatively agree with those of
\citet{Bhardwaj:2023avo}, it should be noted that the values of ${\rm
DM}_{\rm src}$ derived in our model are significantly smaller. While
they obtain $M_{\rm ej} \lesssim 10^{-8}\,M_{\odot}$ to satisfy the
observed DM constraint, we infer $M_{\rm ej/w} \lesssim
10^{-7}\,M_{\odot}$ which is less restrictive. This is due to the fact
that \citet{Bhardwaj:2023avo} assume the full ionization of the ejecta
nuclei in their analysis, which significantly overestimates the number
of free electrons, and therefore their integrated column density. In our
model, we find that $1/\mu_e\sim(2-3)\bar{A}^{-1}$ for both the
dynamical ejecta and disk wind, which is a more realistic assumption. In
addition, \citet{Bhardwaj:2023avo} obtain $M_{\rm ej} \lesssim
7\times10^{-15}\,M_{\odot}$ for the radio signal to be not attenuated
due to free-free absorption. From our model, we have a less stringent
limit $M_{\rm ej/w} \lesssim 10^{-8}\,M_{\odot}$ for the optical depth
of free-free absorption to be less than unity.

\section{Conclusions}
\label{sec:conclusions}
Motivated by the potential association between GW190425 and FRB
20190425A \citep{Moroianu:2022ccs}, we have considered a scenario in
which the \ac{NS} merger remnant survived over secular time scale. We
have performed numerical relativity simulations targeted to GW190425
with the BA \ac{EOS} \citep{Fattoyev:2020cws}, which satisfies
most presently available astronomical and laboratory constraints on the
dense matter \ac{EOS} and predicts a maximum mass for a nonrotating
\ac{NS} of $2.6\ \Mo$. The chirp mass for all simulations has been
fixed to $\mathcal{M} = 1.44\ \Mo$, and we have explored four mass ratios:
$q = 1, 1.17, 1.33$ and $1.67$. With the exception of the $q = 1.67$
binary, all other binaries produce secularly stable remnants.

We have found that all binaries produce massive outflows with $M_{\rm
ej} \lesssim 0.01\ \Mo$ through a combination of tidal torques and
shocks at merger. Even the $q = 1.67$ binary, which undergoes prompt
\ac{BH} formation, produces a massive outflow as a result of the tidal
disruption of the secondary star. The dynamics for this latter binary
are similar to that discussed in \citet{Bernuzzi:2020txg} for a lower
mass system with a softer \ac{EOS}. In addition to the dynamical ejecta,
all systems produce remnants with massive accretion disks ($M_{\rm
disk} = 0.1\ \Mo - 0.3\ \Mo$). It is expected that $20{-}40\%$ of the
disk will also be unbound on secular timescale, further contributing to
the mass ejection \citep{Shibata:2019wef}.

Due to the large mass ejection, our models produce bright kilonovae. At
the distance inferred for FRB 20190425A, the kilonova is expected to
peak at 20~mag in both $g$ and $r$ bands within 1~day after merger, for
the $q = 1, 1.17$ and $1.33$ binaries, which form long-lived remnants.
There are several stringent upper limits from ZTF, GOTO, and CSS at the
sky position of FRB 20190425A in the first two days. The three $r$ band
ZTF upper limits of 21~mag in the first day post-merger are particularly
constraining and disfavor a scenario in which FRB 20190425A is triggered
by the delayed collapse of the merger remnant.  The formation of a
stable remnant, and hence the BA \ac{EOS}, is disfavored even if we
discard the association with FRB 20190425A, due to the non detection of
any kilonova despite the extensive search \citep{Coughlin:2019xfb}.
However, it is important to keep in mind that not all the plausible sky
locations for GW190425 have been observed and our kilonova model
possibly suffers from systematic uncertainties difficult to quantify with
accuracy.  Therefore it is not possible to definitively exclude the
formation of a long-lived remnant for GW190425 from the lack of a robust
kilonova identification. The kilonova would also have escaped detection
if the merger was at more than ${\sim}150\ {\rm Mpc}$, which is not
unlikely given that the inferred distance for GW190425 from \ac{GW} data
alone was $159^{+69}_{-71}\ {\rm Mpc}$ \citep{Abbott:2020uma}.

The presence of a dense shell of ejecta between us and the compact
object formed in GW190425 also affects the propagation of a possible
radio burst generated by the remnant.  The requirement that the medium
should be transparent to radio waves at the relevant frequencies
translates to a very stringent constraint on the ejecta mass. The
observed DM of the FRB poses an additional constraint. We have revised
the estimates of \citet{Bhardwaj:2023avo} using the data from our
simulations. We have also taken into account the fact that heavy
elements have a very large ionization barrier, so they are not fully
ionized, but only singly or doubly ionized at the relevant times.
Accordingly, we find significantly smaller DMs and absorption opacities.
While \citet{Bhardwaj:2023avo} placed an upper limit on the ejecta mass
of $7\times 10^{-15}\ \Mo$, the constraint that we obtain are much
weaker: $M_{\rm ej} \lesssim 10^{-8}\ \Mo$. However, the dynamical
ejecta mass from the simulations is at least five orders of magnitude
larger. As discussed previously, the secular disk wind will contribute
even more material, likely increasing the overall mass of the outflows
to ${\sim}10^{-1}\ \Mo$. As such, our conclusions agree with those of
\citet{Bhardwaj:2023avo}: FRB 20190425A and GW190425 are not associated.

The DM contribution of the ejecta evolves with time as ${\rm DM}_{\rm
src} \propto t^{-2}$, so an FRB with a DM comparable to that of FRB
20190425A, could only have been produced with a delay of ${\sim}
18\,{\rm days}$, taking into account just the dynamical ejecta. With the
inclusion of the disk wind contribution to DM, this time increases to
${\sim} 173\,{\rm days}$. Similarly, for the material to become
optically thin, the dynamical ejecta (disk wind) needs to evolve for
${\sim} 12.4\,(63)\,{\rm days}$. As such, FRBs from possible stable
\ac{NS} merger remnants should only be expected with a delay of a few
months from the merger. This picture could be modified for events
observed close to on-axis, because the \ac{SGRB} could clear the region
at high-latitude of ejecta, thereby lowering the opacity and DM 
\citep{Zhang:2013lta, Wang:2023yqy}. On the other hand, this argument
would not apply to GW190425, because the inferred observation angle
assuming an association to FRB 20190425A would be larger than $40^\circ$
with high confidence \citep{Bhardwaj:2023avo}.

This study is a cautionary tale that underscores our incomplete
understanding of \ac{NS} mergers and their dynamics. Contrary to the
widespread notion that GW190425 was a prompt collapse event, we have
shown that the formation of a long-lived remnant cannot be excluded on
the basis of our current knowledge of the \ac{NS} \ac{EOS}. We can
confidently exclude that GW190425 was associated with FRB 20190425A, but
we cannot exclude the formation of a long-lived remnant, due to the
incomplete coverage of the relevant sky regions and the faintness of the
potential optical/infrared counterpart. Future observations of
GW190425-like events, or deeper upper limits, can constrain models, like
the BA \ac{EOS}, with ``large'' maximum \ac{NS} mass. This would have a
profound impact on our understanding of the properties of matter at the
most extreme densities realized inside \acp{NS}. Perhaps, the most
important lesson from this work is that the allocation of resources for
electromagnetic follow up of \ac{NS} mergers should be informed by the
properties of the source, such as their chirp mass, rather than by
theoretical prejudiced metrics. We urge the LIGO-Virgo-KAGRA
collaboration to publicly release the chirp mass of future \ac{NS}
merger events promptly.

\section*{Data Availability}
Data generated for this study will be made available upon reasonable
request to the corresponding authors.

\section*{Acknowledgements}
The authors thank Mattia Bulla, Antonella Palmese, Bing Zhang and
Eleonora Loffredo for useful discussions.
DR acknowledges funding from the U.S. Department of Energy, Office of
Science, Division of Nuclear Physics under Award Number(s) DE-SC0021177
and from the National Science Foundation (NSF) under Grants No. PHY-2011725,
PHY-2020275, PHY-2116686, and AST-2108467.
GR acknowledges support by the Deutsche Forschungsgemeinschaft (DFG,
German Research Foundation) – Project-ID 279384907 – SFB 1245.
GR acknowledges support by the State of Hesse within the Research
Cluster ELEMENTS (Project ID 500/10.006).
M.B. acknowledges support from the Eberly Postdoctoral Research Fellowship at
the Pennsylvania State University.
The work of A.P. is partially funded by the European Union under
NextGenerationEU. PRIN 2022 Prot. n. 2022KX2Z3B.
K.M. is supported by the NSF Grant No.~AST-1908689, No.~AST-2108466 and
No.~AST-2108467, and KAKENHI No.~20H01901 and No.~20H05852.
This research used resources of the National Energy Research Scientific
Computing Center, a DOE Office of Science User Facility supported by the
Office of Science of the U.S.~Department of Energy under Contract
No.~DE-AC02-05CH11231. Computations for this research were also
performed on the Pennsylvania State University's Institute for
Computational and Data Sciences' Roar supercomputer.


\bibliographystyle{mnras}




\acrodef{ADM}{Arnowitt-Deser-Misner}
\acrodef{AMR}{adaptive mesh-refinement}
\acrodef{BH}{black hole}
\acrodef{BBH}{binary black-hole}
\acrodef{BHNS}{black-hole neutron-star}
\acrodef{BNS}{binary neutron star}
\acrodef{CCSN}{core-collapse supernova}
\acrodefplural{CCSN}[CCSNe]{core-collapse supernovae}
\acrodef{CMA}{consistent multi-fluid advection}
\acrodef{CFL}{Courant-Friedrichs-Lewy}
\acrodef{DG}{discontinuous Galerkin}
\acrodef{HMNS}{hypermassive neutron star}
\acrodef{EM}{electromagnetic}
\acrodef{ET}{Einstein Telescope}
\acrodef{EOB}{effective-one-body}
\acrodef{EOS}{equation of state}
\acrodef{FF}{fitting factor}
\acrodef{GR}{general-relativistic}
\acrodef{GRLES}{general-relativistic large-eddy simulation}
\acrodef{GRHD}{general-relativistic hydrodynamics}
\acrodef{GRMHD}{general-relativistic magnetohydrodynamics}
\acrodef{GW}{gravitational wave}
\acrodef{ILES}{implicit large-eddy simulations}
\acrodef{LIA}{linear interaction analysis}
\acrodef{LES}{large-eddy simulation}
\acrodefplural{LES}[LES]{large-eddy simulations}
\acrodef{MRI}{magnetorotational instability}
\acrodef{NR}{numerical relativity}
\acrodef{NS}{neutron star}
\acrodef{PNS}{protoneutron star}
\acrodef{RMNS}{remnant massive neutron star}
\acrodef{SASI}{standing accretion shock instability}
\acrodef{SGRB}{short $\gamma$-ray burst}
\acrodef{SPH}{smoothed particle hydrodynamics}
\acrodef{SN}{supernova}
\acrodefplural{SN}[SNe]{supernovae}
\acrodef{SNR}{signal-to-noise ratio}
\acrodef{ZAMS}{zero age main sequence}

\bsp	
\label{lastpage}
\end{document}